# Photonic Physical Unclonable Functions: From the Concept to Fully Functional Device Operating in the Field

M. Akriotou*[a], A. Fragkos[b], and D. Syvridis[a]

[a] Department of Informatics & Telecommunications, National and Kapodistrian University of Athens, Panepistimiopolis Ilisia 15784, Athens, Greece; [b] Eulambia Advanced Technologies Ltd. Ag. Ioannou 24, 15342, Athens, Greece;

## ABSTRACT

The scope of this paper is to demonstrate a fully working and compact photonic Physical Unclonable Function (PUF) device capable of operating in real life scenarios as an authentication mechanism and random number generator. For this purpose, an extensive experimental investigation of a Polymer Optical Fiber (POF) and a diffuser as PUF tokens is performed and the most significant properties are evaluated using the proper mathematical tools. Two different software algorithms, the Random Binary Method (RBM) and Singular Value Decomposition (SVD), were tested for optimized key extraction and error correction codes have been incorporated for enhancing key reproducibility. By taking into consideration the limitations and overall performance derived by the experimental evaluation of the system, the designing details towards the implementation of a miniaturized, energy efficient and low-cost device are extensively discussed. The performance of the final device is thoroughly evaluated, demonstrating a long-term stability of 1 week, an operating temperature range of $5^0$C, an exponentially large pool of unique Challenge-Response Pairs (CRPs), recovery after power failure and capability of generating NIST compliant true random numbers.

## 1. INTRODUCTION

The modern world is an interconnected one. The trend of "everything connected" leads to a landscape of cloud computing, mobile devices and the ubiquitous "Internet of Things". That, however, also leads to a vastly enlarged attack surface, vulnerable to all kinds of cyber-attacks. At the same time, data generation is continuing its exponential increase. At present, humanity is producing around 2.5 quintillion bytes per day [1] and it is certain that that number will double in the next couple of years. In order to safeguard all that data, new ways have to be explored for securing the devices that produce, process or store them, as well as their interconnection. Current authentication protocols based simply on storing keys cannot be trusted since these keys can be stealth stolen or leaked, and knowledge-based authentication methods like passwords or PINs have other weaknesses which are actively being exploited. This "security deficit" will only get worse, as IoT devices are being deployed at ever-increasing numbers. This deficit could be mitigated by previously under-explored technologies like Physical Unclonable Functions (PUFs).

PUFs are the physical analogues of one-way mathematical functions, which operate on the principle of capitalizing on the inherent randomness of physical objects. They can generate unpredictable but reproducible outputs (responses) when external stimuli are applied (challenges) [2]. Their primary advantage is their immunity to physical cloning and reverse-engineering which, combined with their deterministic operation, makes them ideal for on-demand cryptographic key generation [3], [4], thereby making obsolete any conventional methods of stored keys. Other uses may include providing secure links between software and hardware [5], used as unique tokens for authentication [6], [7], and anti-code-reuse schemes [8].

The currently prevalent PUF design is the electronic PUF [9]–[14] which has been making inroads, despite having been proven to be vulnerable to machine learning [15]–[17] and side-channel [18]–[20] attacks. Optical PUFs, on the other hand, have largely remained in obscurity, however promising they may be, something which could be attributed to their challenging integration with electronic devices.

* makriotou@di.uoa.gr

Optical PUFs were introduced in [2]; their principle of operation based on the random interference patterns (speckles) produced when a laser beam propagates through an inhomogeneous material. Proposed implementations include transparent tokens containing randomly distributed micro-structures [2], [21], laser-engraved elements [22], [23] or even sheets of paper [24]. The general principle of operation of an optical PUF involves a coherent light source (laser) illuminating the surface of the element, while the reflection or transmission of the light constitutes the complex response (speckle pattern) which is digitized and processed; this can be used either as a unique "fingerprint" [25] or as an entropy source for random number generation [26]. Varying different parameters of the coherent light source leads to the alteration of the illumination conditions, thereby providing different "challenges" [27]; the wavelength, the number of beams or their diameter, the angle of incidence, or the specific spatial intensity are a few examples.

Having identified the underutilization of optical PUFs, we decided to investigate the feasibility of creating such a device, employing standard diffusing optical elements as well as unpolished polymer optical fibers (POFs) as tokens, evaluating their performance by using two of the aforementioned illumination parameters, namely the wavelength and the spatial intensity distribution. Standard metrics, common throughout current literature, were used in order to quantify the main security features of the devices in terms of robustness, unclonability and unpredictability. For this purpose, we designed various experimental setups as well as a specific protocol for processing the data and producing cryptographically compatible binary sequences.

## 2. PHOTONIC PUF

### 2.1 Hardware

During the experimental evaluation of the photonic PUFs, two different setups were primarily used. The first setup comprised of a CW external-cavity laser source, amplified by an Erbium Doped Fiber Amplifier (EDFA), guided through a single mode fiber (SMF) to the token, after which the resulting speckle pattern was collected by a positive lens of 6mm focal length and finally captured by a vidicon camera with a resolution of 340x340 pixels.

The second setup utilized a HeNe laser source, emitting at 652nm, projected through a lens on an 8-pixel grid of an LCD screen, after which followed a lens and the token. The output of the token was captured by a conventional web camera with a resolution of 1280x960 pixels.

The first setup was used as a proof-of-operation for the photonic PUF and for further evaluation of the performance of the PUFs when the wavelength is used as the challenge and the rest of the illumination parameters remain constant. The second setup was used to further study the performance of the PUFs for different combinations of switched on and switched off pixels of the LCD, corresponding to $2^8$ different spatial intensity distributions of the incident beam.

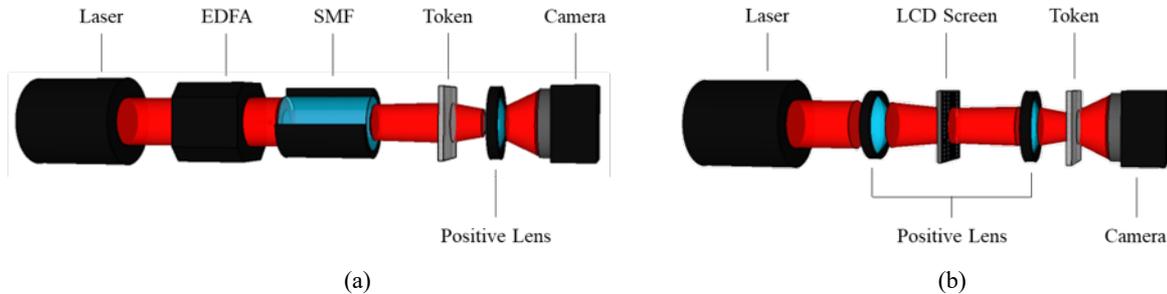

Figure 1: the two experimental setups used for the evaluation of the PUF performance when a) the wavelength is used as the challenge and b) spatial distribution used as challenge. The token in both cases corresponds to either a conventional optical diffuser or an unpolished POF

### 2.2 Protocol

The usefulness of any PUF system relies on the deterministic behavior of its physical component, whereas noise in any form will invariably disrupt the reproducibility of its responses. Many techniques [28] have been explored and utilized for the elimination of the effects of noise, including signal processing, hashing, fuzzy extractors [29], alongside ECCs (error correcting codes). A number of these techniques were incorporated in a general security protocol [30] for our system, selected to optimally conform with its most important properties. Figure 2 illustrates the protocol as well as the primary parts of the algorithm developed to generate the necessary binary sequences from the captured raw data.

In short, all the different responses produced by applying the same challenge x on a component p, are mapped to a unique binary key K, based on a fixed set of ECC parameters. The procedure is always carried out in two different modes; the enrollment mode and the authentication mode. The former corresponds to the first time that the challenge is applied, whereby the output K is generated, along with a set of appropriate helper data h. The latter represents the rerun of the measurement, during which the attempt to recreate the same result K is made, by using the helper data produced in the enrollment phase. At the same time, the ECC algorithm is utilized for the detection and the correction of all the possible K results' discrepancies, leading to the decrease of their in-between bit error rate.

The first schematic of Figure 2 represents the enrollment mode of the system, during which a challenge x is applied for the first time. Initially, the PUF's response is processed via a hashing algorithm and a binary sequence of length M is extracted, while all the necessary information to reproduce it, is stored in a data file $h_H$. This sequence is the output key $K_E$ of the system. Simultaneously, a random binary secret of length $\ell \leq M$ is independently generated and subsequently encoded through a BCH algorithm. Then, a second helper data file $h_K$ is stored, containing the result of the XOR operation between the binary sequence $K_E$ and the encoded random secret.

The second schematic of Figure 2 illustrates the authentication mode of the system; a response is obtained under the same challenge as in the enrollment phase. The corresponding hashed binary sequence $K_A'$ is calculated through the helper data $h_H$, and the XOR between $K_A'$ and $h_H$ helper data is extracted, resulting in the error-affected encoded secret. The latter is subsequently decoded employing the inverse BCH code and encoded again. This encoded result is then XORed with the helper data $h_K$, which is the final output of the system $K_A$.

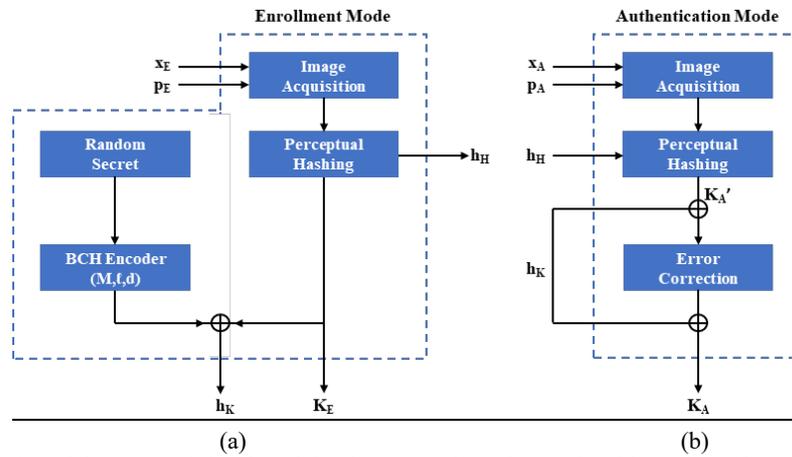

Figure 2: the two modes of the protocol developed for the generation of error-free binary keys from the noisy responses of the PUF.

It is noteworthy that the selected BCH parameters (M, $\ell$, d) play a significant role upon the effectiveness of the procedure described above. As the ECC theory dictates, when $\ell$ is the number of information bits encrypted in a code - message of length $M = 2^k - 1$, and d corresponds to the minimum number of bits that differ between two codes, the error correction capability of a BCH algorithm is given by the relation $t = (d - 1) / 2$. Essentially, the employed algorithm can detect and correct up to t errors through the redundant bits added during the encoding process of the enrollment mode. The length $\ell$ has an inversely proportional relation to the available overhead and the t capability of the code; given a fixed value of M, increasing $\ell$ results in the correction of fewer bit-flip-errors, and subsequently leads to the reduction of the responses' number that is ultimately mapped to one single output K.

With respect to the protocol described in the previous paragraphs, the three main properties [30] of PUFs can be formally defined and quantified through the following equations:

$$\text{Robustness} := \Pr[\text{PUF}(p, x, h) \to (K, h): \text{PUF}(p, x, \varepsilon) \to (K, h)] \qquad (1)$$

, where x is an applied challenge to one single component p and K its corresponding output. The $\varepsilon$, h symbols refer to the input helper-data files of the enrollment mode and the authentication mode respectively, with $\varepsilon$ being empty. In other words, robustness for a given system expresses the conditional probability of generating an identical output K in both modes using the helper data created in the enrollment.

$$\text{Unclonability} := \Pr[\text{PUF}(p', x, h) \to (K, h): \text{PUF}(p, x, \varepsilon) \to (K, h)] \quad (2)$$

On the other hand, physical unclonability (or probability of physical cloning) is substantially defined as the probability of generating an identical output K by two different systems, via the helper data h created in the enrollment mode of the latter. These two systems are characterized by the same experimental conditions, while their corresponding physical components differ (i.e. $p' \neq p$). Finally, physical unpredictability is defined as the conditional probability of generating an identical output K by applying two different challenges x', x on the same component p, via the helper data created in the enrollment mode of the latter.

$$\text{Unpredictability} := \Pr[\text{PUF}(p, x', h) \to (K, h): \text{PUF}(p, x, \varepsilon) \to (K, h)] \quad (3)$$

Generally, the key extraction process, as described in the previous paragraphs, can be put carried out for the majority of the existing PUF implementations without the hashing step. In our case however, the nature of the produced responses, along with the fuzzy extractor's requirement for a binary output, necessitates its introduction. Two different techniques of image hashing were investigated in order to detect which one offers the optimal performance of our PUF with the minimum computational cost.

### 2.3 Hashing Techniques

As previously mentioned, the experimental evaluation of photonic PUFs entails the use of an imaging device, resulting in an unprocessed image for each run. Therefore, it was necessary to identify the most suitable image hashing techniques for the PUF's purpose. The first such technique was Singular Value Decomposition (SVD), which is used to extract the invariant geometric features of an image and which produces robust, content-based hashes. SVD is an algebraic transformation stating that every real matrix $A \in \mathbb{R}^{M \times N}$ can be factorized in the form of $U \cdot \Sigma \cdot V^T$, where $\Sigma \in \mathbb{R}^{M \times N}$ is a rectangular diagonal matrix that contains the square roots of the non-zero eigenvalues of $A^T A$ (i.e. singular values), positioned in descending order. $U \in \mathbb{R}^{M \times M}$ and $V \in \mathbb{R}^{N \times N}$ are two square unitary matrices that contain a set of orthonormal eigenvectors (i.e. singular vectors) of $AA^T$ and $A^T A$ matrices respectively, following the sorted order of their corresponding eigenvalues. The rank of A equals the number of non-zero diagonal elements in $\Sigma$. In most applications, the singular values (SVs) of a matrix decrease quickly with increasing rank, while smaller SVs being more affected by noise [31]. In our case, where matrix A corresponds to the pixel intensities of an image, its singular values (SVs) and vectors (SCs) carry information about the luminance and the geometry of the image respectively.

In that context, SVD can allow the reduction of noise as well as the compression of our responses by keeping the SCs that correspond only to the higher values of their SVs. The algorithm employed for that purpose, which was based on [32] contains the following steps:

- The image, obtained with $N_1 \times N_2$ pixel resolution, is standardized and divided into p overlapping blocks of $k_1 \times k_1$ size, on each of which SVD is applied: $A_i = U_i \Sigma_i V^T_i$ with $1 \leq i \leq p$
- The first eigenvectors of all the U and V matrices are kept and concatenated to create an intermediate image $\Gamma = [u_1, u_2, \ldots, u_p, v_1, v_2, \ldots v_p]$
- The intermediate image $\Gamma$ is then pseudo-randomly divided into r overlapping blocks of $k_2 \times k_2$ size, on each of which SVD is applied: $B_i = U_i \Sigma_i V^T_i$ with $1 \leq i \leq r$
- Again the first eigenvectors of all the U and V matrices are kept and concatenated to create a hash $h = [u_1, u_2, \ldots, u_r, v_1, v_2, \ldots v_r]$
- Every element of h is quantized according to the following rule: an element becomes zero if its value is lower than the value of the element on its right
- Lastly, M elements are pseudo-randomly chosen to constitute the final result

With regard to the described security protocol, the helper data file $h_H$ that is created during enrollment contains the starting indices of the blocks for both SVDs, as well as the indices of the M elements constituting the final hash.

Another hashing technique explored was the Random Binary Method (RBM), which produces content-independent results and is considered non-adaptable. RBM works by capitalizing on the inherent sparsity of signal, enabling its reconstruction even from a lower sampling rate than the Shannon-Nyquist theorem would demand. This is achieved by acquiring a few incoherent measurements directly from the signal by multiplying it with a random table $\Theta$. The entries of this matrix, namely the sensing matrix, are usually selected from an identical & independent distribution (i.i.d.) e.g. Gaussian, sub-Gaussian or Bernoulli [33]. The method can be summarized through the following relation:

$$\tilde{y} = \text{sign}(\Theta y) \quad (4)$$

, where y is the standardized PUF response converted to a one - dimensional array of size $N = N_1 \times N_2$ $\tilde{y}$ is its corresponding resultant hashed bit-string of length $M \leq N$ and sign the quantization function of the procedure.

In practice, the sensing matrix used was equal to $\Theta = S \cdot F \cdot U$. U represents a diagonal random table (NxN), containing only the values $\pm 1$ with Pr $[U_{ii} = 1]$ = Pr $[U_{ii} = -1] = 0.5$, and F the discrete Fourier table of NxN dimensions. Finally, S symbolizes a matrix containing M entries randomly chosen from a uniform distribution (0, N); those were the indices of the (FU)·y elements, being extracted to constitute the corresponding hashed binary sequence $\tilde{y}$. The result of the above procedure is converted to binary by thresholding the real part of each item by the mean. This alternative algorithm, as proposed in [23] has been proven to manifest a similar behavior to the conventional technique with a full Gaussian or sub-Gaussian distribution, providing however a faster extraction of the hashes.

With regard to the described security protocol, in this case, the helper data file $h_H$ that is created during enrollment, contains the three tables U, F and S that constitute the sensing matrix $\Theta$.

## 2.4 Evaluation Metrics

For the valid comparison of the results and for the complete evaluation of the system's performance, apart from the aforementioned probabilities of the definitions 1-3, the following additional metrics were employed:

- The Euclidean Distance (ED) between two vectors $\vec{v}, \vec{u} \in \mathbb{R}^N$ is mathematically defined as:

$$ED(\vec{v}, \vec{u}) = \|\vec{v} - \vec{u}\| = \sqrt{\sum_{i=1}^{N}(v_i - u_i)^2} \quad (5)$$

and is used to determine the extent of the pixel intensity variations between the obtained speckle images.

- The Hamming distance (HD), is the number of bits that differ between two binary sequences s, t $\in \{0,1\}^M$ and it is given by the following relation:

$$HD(s, t) = \sum_{i=1}^{M} v_i \oplus u_i \quad (6)$$

The derived distributions from these two metrics are visualized in the form of histograms usually plotted in pairs; robustness - unclonability and robustness - unpredictability respectively. The main reason for these pairs being the verification of potential overlaps between them. If such an overlap exists, we run the risk of an output, which is obtained by a different challenge or component, to contain such a number of discrepancies which would place it within the overlapping range, thereby allowing the BCH routines to inadvertently rectify it and, thus, cause its wrongful registration.

- The Cross-Correlation Coefficient (CC) between two matrices A, B $\in \mathbb{R}^{N \times M}$ given by:

$$CC(A, B) = \frac{\sum_N \sum_M (A_{mn} - \bar{A})(B_{mn} - \bar{B})}{\sqrt{\left[\sum_N \sum_M (A_{mn} - \bar{A})^2\right]\left[\sum_N \sum_M (B_{mn} - \bar{B})^2\right]}} \quad (7)$$

the computation of which can provide us with information regarding the structural similarity between the acquired speckle images.

## 3. RESULTS

### 3.1 Wavelength as a Challenge

The setup of Figure 1a was used in order to evaluate two different tokens for the construction of an optical PUF: an unpolished POF of 980μm core, 20μm cladding and 10cm length, and a conventional optical diffuser. By employing the former, a preliminary evaluation of the system was performed and the optimal wavelength step between two consecutive challenges was detected. The 100pm increment was chosen, since it provides a sufficiently low cross-correlation coefficient between two successive speckle patterns thus the challenges produce adequately discernible responses [26]. The challenge space that was eventually chosen is the result of varying the wavelength of the laser between 1540 and 1570nm at 100pm intervals.

Then, 300 unpredictability images were captured from a single diffuser and a single fiber over the wavelength range mentioned above. The cross-correlation coefficients of the two datasets calculated by having the 1540nm image as reference, are presented in Figure 3. The results showed that the POFs are very sensitive to wavelength variations, whereas the diffuser exhibits a lower rate of correlation decrease, indicating that different wavelengths do not result in adequately dissimilar responses, thereby decreasing the challenge space dramatically.

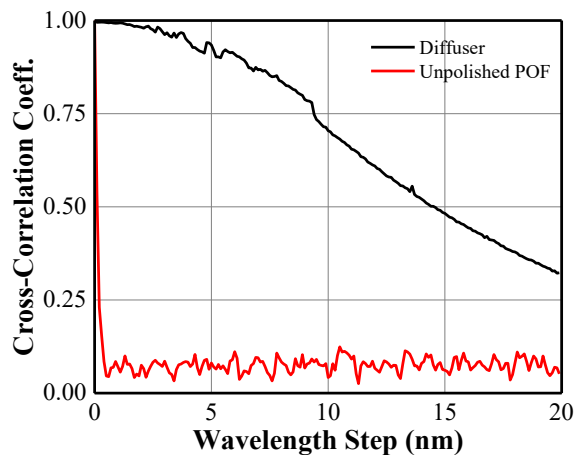

Figure 3 Cross-Correlation coefficient of an image acquired under 1540nm excitation and its subsequent responses produced by increasing the laser wavelength in 100pm step.

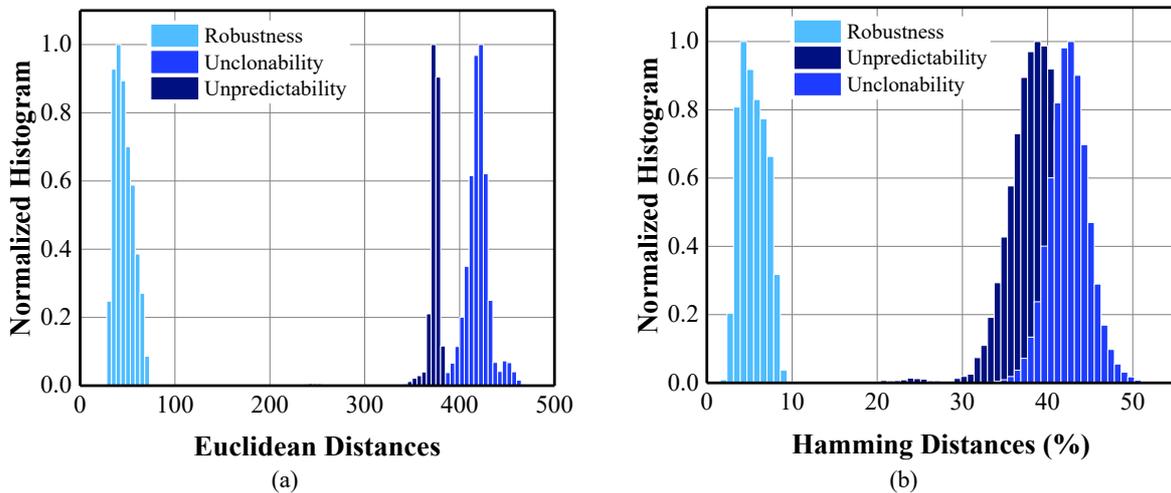

Figure 4: a) Normalized histogram of the Euclidean distances between 60 robustness images, 1000 unclonability images and 300 unpredictability images. b) Hamming distances between the hashed binary sequences of the same datasets extracted via the RBM method

Therefore, POFs are demonstrably superior to diffusers for this method of challenge generation, something which led to their further study in greater detail. Three sets of measurements were performed, since the system evaluation metric is trifold: unpredictability, robustness and unclonability. Initially, 300 images were recorded from keeping a single fiber and varying the wavelength of the laser from 1540nm up to 1570nm, at 100pm intervals as usual. The magnitude of variation within these images quantifies the unpredictability. Secondly, keeping constant the waveguide and illumination, 60 images were recorded over a substantial interval (multiple minutes), the noise immunity of the system was quantified as robustness. Finally, keeping constant illumination and applying it to 1000 different waveguides with distinct facet microstructures, the unclonability of the system was similarly identified. Those three datasets were used for computing the Euclidean distances between images of each set; the results can be seen in Figure 4a. It is clear that the robustness distribution is clearly distinct from the unclonability and unpredictability distributions, with absolutely no overlap between them, something which, as already mentioned, is required in the operation of such systems. This requirement ensures that no two challenges (illumination conditions) could ever be misconstrued as the same challenge affected by noise, thereby producing a false positive, or no two takes of the same challenge to be mistaken for different challenges, producing a false negative. Similarly, no two waveguides could be misconstrued as the same waveguide and no two takes of the same waveguide be mistaken as different waveguides.

Figure 4b shows the percentile Hamming distance distributions as calculated for the binary sequences produced by the three aforementioned datasets' images. Those sequences were extracted by using the RBM technique, keeping the helper data constant. The Hamming distances were calculated prior to the application of the BCH. It is evident that the distributions are following the same patterns as that of the Euclidean distances. The average and standard deviation for these three datasets are $\mu_{noise}$ = 5.4±1.5, $\mu_{wave}$ = 38.6±3.4 and $\mu_{fiber}$ = 42.8±1.9, where, ideally, they should be $\mu_{noise}$ = 0, $\mu_{wave}$ = $\mu_{fiber}$ = 50%. While not far from ideal, the system could be further improved by tweaking the details; for instance, unpredictability could be improved by slightly increasing the wavelength step between challenges.

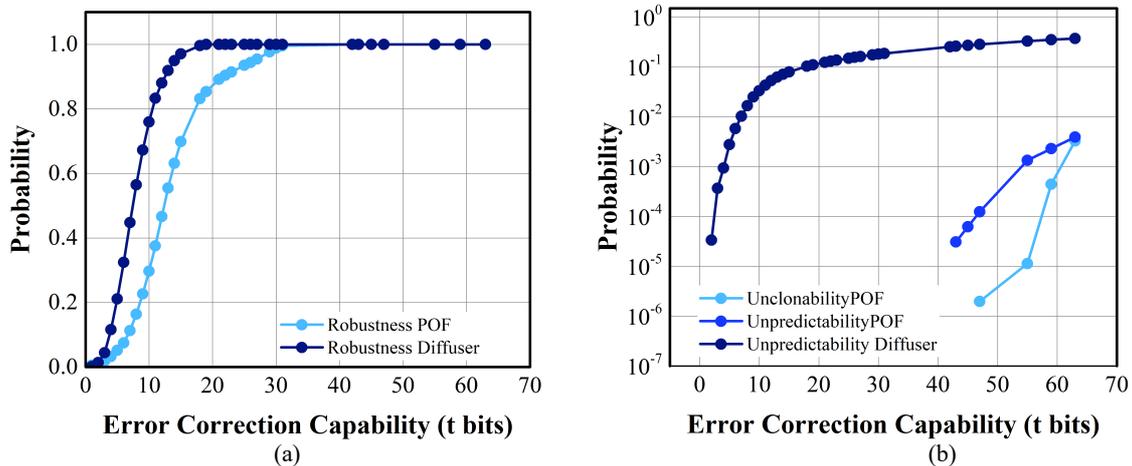

Figure 5 a) Probability of generating identical binary sequences from robustness images by both modes of the protocol, using the helper data created in the enrollment. b) Probability of generating identical binary sequences from unclonability and unpredictability images by both modes of the protocol, using the helper data created in the enrollment. The corresponding results of the diffuser are also included in the graphs.

The next step was to investigate the probability of the protocol producing identical outputs after applying the BCH code [25]. The probabilities were calculated through eq. 1-3 and for every possible error correction capability for a set key length M = 255 bits. Figure 5a shows those probabilities for the robustness image set, along with the corresponding results of the diffuser. It is apparent that the minimum error correction capability required for the probability to be equal to 1(i.e. noise errors are always corrected) is t = 31 bits for our system, whereas the diffuser only needs t = 19 bits to achieve the same result. However, the physical mechanism responsible for that behaviour also augments the response variance of the PUFs, which will be obvious in the unclonability and unpredictability figures.

In Figure 5b we present those probability values, calculated from the unclonability and unpredictability image sets, along with the diffuser results of, showing the probability that the error correction should mistakenly give the same key from two different PUFs (unclonability) or two different challenges (unpredictability). It is apparent that, while the diffuser's probability of error is already at 0.11 for t = 19 bits and reaches 0.261 for t=43, our system is exhibiting a much lower probability of 3.1*10^(-5) for the latter error correction capability. We can safely say, therefore, that while the POF system needs more bits dedicated to error correction due to higher noise levels, its unpredictability rating is better by 4 orders of magnitude. Similarly, the unclonability rating is slightly better, something which could be foreseen looking at the Hamming distances.

Our investigation reached its conclusion by indicating that, when the stimulus is the variation of wavelength, the unpolished POF appears more suitable as the token in a PUF system than a diffuser, since its overall performance is better, despite the fact that it needs more ECC bits to ensure key reproducibility. The POF system also significantly enhances unclonability [25], which is a measure of how resilient the PUF is against physical replication by either honest (unintentional) or malicious (intentional) manufacturers. The system was also tested through the NIST Statistical Test Suite and was proven to be an adequate random number generator [26]. It is shown to outperform most of the electronic PUF systems, while its performance is equal to the best of them, like the SRAM PUF, or the pseudorandom Ziggurat algorithm [26]. The NIST results along with its robustness performance render our system cyber-secure, making the storage of secure keys irrelevant. In further studies, the tunable laser is replaced by a spatial light modulator (SLM), as proposed in [27], something which introduces a much larger challenge space (number of inputs) with comparable or even higher performance.

### 3.2 Spatial Pattern as a Challenge

The setup of Figure 1b was also used for the evaluation of having an SLM (i.e. LCD screen) as the control element. Using an unpolished POF of 3cm in length and the same diffusing element used previously, 256 unpredictability images were captured, with each challenge corresponding to a different beam configuration, produced by a unique pattern of "on" and "off" pixels on the LCD. Another 60 images were captured under identical experimental conditions in the course of several minutes, to evaluate the system's robustness.

Figure 6a and Figure 6b show the Euclidean distances that were calculated for the speckles produced by the fiber and the diffuser. As can be seen, a clear and sizeable tail is present in both unpredictability distributions, which indicates an increased similarity between the speckles, and which would produce closely interrelated and therefore predictable keys. This behavior can be attributed to the fact that every speckle which emerges from 2 or more pixels of the LCD is, in fact, the linear combination of the speckles produced by each pixel separately. Moreover, the fiber exhibits greater susceptibility to mechanical vibrations, something which leads to increased average and standard deviation of its robustness histogram (i.e. $\mu_{noise} = 318.1 \pm 98$ for the fiber while $\mu_{noise} = 91.98 \pm 33.96$ for the diffuser) and, ultimately, to the overlap between the distributions. Practically, this means that images from two different challenges that fall within the overlap could be erroneously mistaken for the same challenge only affected by noise. Conversely, a noisy image could also be mistaken for a different challenge. In effect, this overlap is an indication that the PUF functionality is not within acceptable limits and if used in this manner, would lead to an unacceptable number of false positives and negatives.

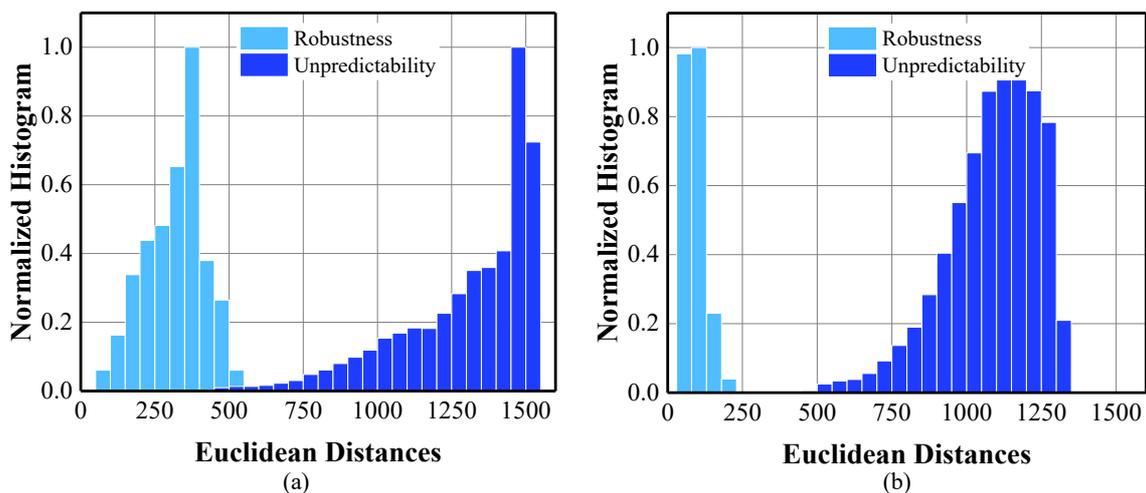

Figure 6 a) Normalized histogram of the Euclidean distances between 60 robustness images and 256 unpredictability images from b) an unpolished POF of 3cm length and b) the diffuser

We explored two alternatives for addressing this deficiency, the first being an attempt to separate the Hamming histograms algorithmically, through finding a better suited hashing method, one that would better allow for and enhance the geometric features of each image. The SVD method was selected and the comparative results are shown in Figure 7, where Figure 7a presents the distributions calculated via the content-agnostic RBM method and which follows the same pattern as that of the Euclidean distance distribution, exhibiting the same overlap. However, the adaptive SVD method lacks any overlap. The downside is that SVD is computationally demanding, making it unsuitable for real-time applications without the use of dedicated hardware.

The second approach for addressing the aforementioned deficiency of the overlap is to revise the way the LCD patterns are generated; rather than use all possible combinations of on/off pixels for a given grid, the number of switched-on pixels was kept constant, trying all different permutations on the grid. With this method, the results of the POF and the diffuser are comparable regarding unpredictability. Therefore, the diffuser can be considered preferable in this case, since it requires fewer optical elements for guiding the beam.

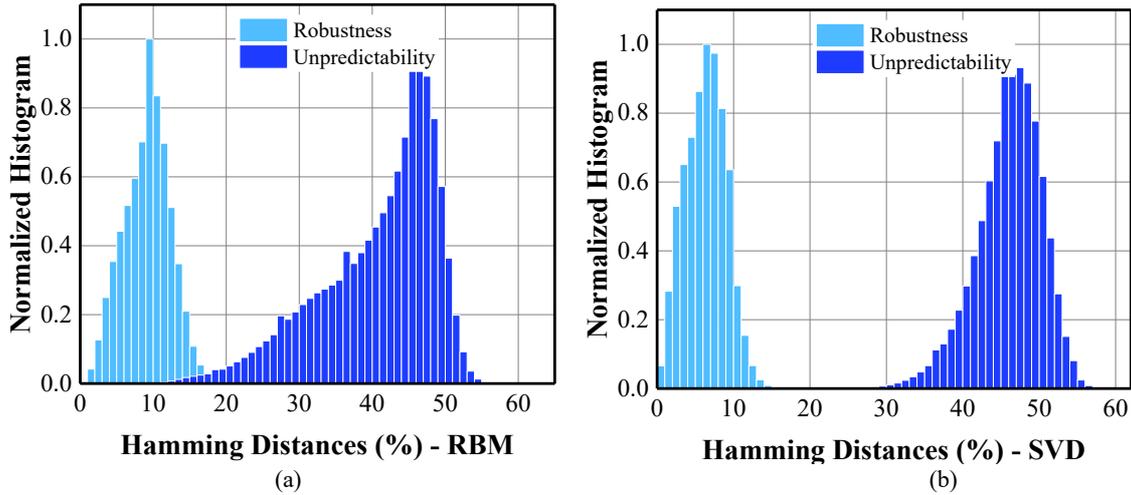

Figure 7 Normalized histograms of the Hamming distances between 60 robustness images and 256 unpredictability images from the POF, extracted via a) the RBM method and b) the SVD method

## 4. PRACTICAL IMPLEMENTATION

In this section a fully working, standalone and miniaturized photonic PUF will be demonstrated based on cost-effective components. In the first paragraph, an extensive description of the component selection, hardware design and overall assembly of the PUF module will be presented, while in the second paragraph we will thoroughly evaluate the performance of the module based on the three main properties of the device, namely robustness, unpredictability and unclonability, using the metrics of the previous analysis.

### 4.1 Hardware design

According to the results demonstrated in section 3, a diffuser used as a PUF token provides increased mechanical stability and lower wavelength dependency compared to a POF. Additionally, the intensity pattern as a challenge mechanism can provide an exponentially larger number of challenges with lower implementation cost and increased challenge reproducibility compared to the wavelength-based challenges. Taking into consideration the aforementioned advantages, a photonic PUF based on diffuser using intensity pattern as challenge generator has been developed and details regarding the specific design will be discussed in the next paragraph.

### 4.2 Component Selection

The implementation of a standalone photonic PUF module is based mainly on the building blocks incorporated in the experimental setup demonstrated in section 2.1. In the following paragraphs we breakdown the list of the necessary components and we discuss the criteria on which we have based our selection.

Laser source: This component is one of the most crucial parts of the device as it provides the stimulus for the PUF token. The significant requirements for this component are the wavelength stability, power consumption, output optical power, footprint and cost. Therefore, a low-cost semiconductor laser was selected, with a central emission peak at 635nm and adjustable output optical power up to 5mW, appropriately stabilized in terms of temperature and current. The wavelength selection was realized taking into consideration the wide range of commercially available cameras at the visible spectrum.

Intensity Pattern Generator: As mentioned in section 3.2, a Spatial Light Modulator such as an LCD can be used in order to generate different patterns. The significant requirements for this component are the resolution, and more specifically the pixel density, refresh rate, contrast, footprint and cost. The module that is selected has a resolution of 640x360 pixel in a form factor of 0.2" and a refresh rate of 33kHz.

PUF token: For the specific implementation a diffuser element has been chosen as the PUF token, due to the advantages mentioned in the previous section. The significant requirements for this component are the size of the random scatterers at the surface of the glass and the coefficient of thermal expansion (CTE). Large number of scatterers with adequately small size is preferable because it increases the complexity of the generated speckle pattern and hinders the modelling attacks based on ray-tracing methods. In this category the most suitable material is the fused silica glass which offers a CTE equal to $0.55 \times 10^{-6}/^0C$. In the market there is a wide range of grit sizes which produce different grain size. We selected 1500 grit which provides the finest and more evenly distributed scatter grain.

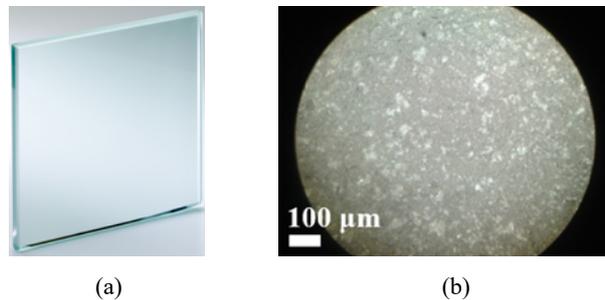

(a)           (b)
Figure 8. Fused silica diffuser panel (a) and microstructure (b)

Camera module: In order to capture the speckle pattern images that are generated by the PUF token, a camera module must be incorporated. The most significant requirements for this component are the resolution, the frame rate, the capability of providing adjustment in the image capturing properties, the acquisition of uncompressed images, the footprint and cost. The resolution of the camera is almost the most crucial property because it defines the minimum distance from the output facet of the token where the camera will be placed, and therefore specifies the maximum length of the optical path and the size of the device in extent. The selected module is a 5MP CMOS sensor with 60 frames per second, adjustable image acquisition properties and uncompressed image.

Main Control Unit: This is the heart of the system as it performs all the necessary tasks for the PUF to be fully functional. These tasks are the control of the optoelectronic peripherals, the digital processing of the speckle pattern images for the extraction of the authentication keys and random numbers, the establishment of communication channel with external devices and general housekeeping. The significant requirements for this component are the processing power, energy consumption, footprint and cost. The unit used incorporates a quad core ARM processor at 1.4GHz, which provides adequate processing power for undertaking the aforementioned tasks.

## 4.3 Hardware Assembly

In order to deliver a miniaturized and fully functional device the aforementioned components have to be integrated in a unified and robust setup and harmonically interact with one another. The key points of the design which are crucial for the performance of the system are the robust mechanical assembly, the overall thermal stability and the development of optimized firmware and software.

This module is integrated inside an aluminum enclosure with internal thermal insulation for limiting the ambient temperature variations to interfere with the system. Two fans with adjustable air flow are installed on either side of the case, with one of them exporting the warm air generated from the device and the other importing cold air regulating the internal temperature of the system and stabilizing laser operation. The internal structure of the case with all the supporting components along with the overall assembly of the standalone device is depicted in the figure below. The communication between the system and external devices such as computer terminals or servers is realized using an ethernet connection through an RJ-45 port located at the back of the device.

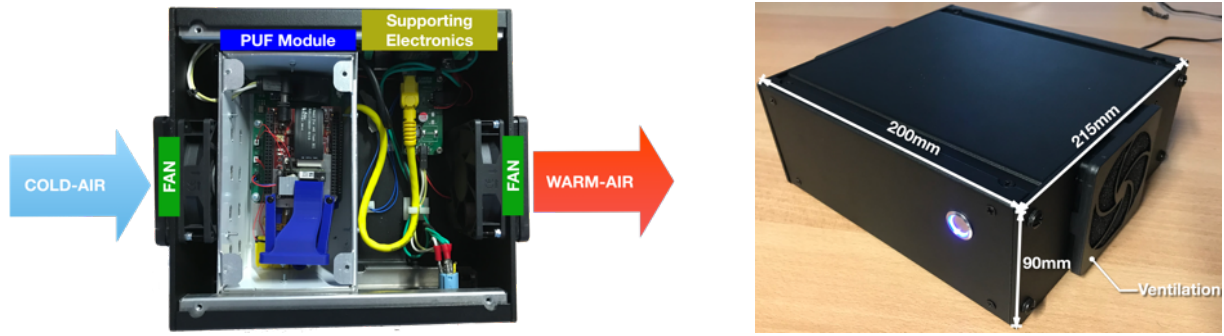

Figure 9. Standalone PUF device

From the firmware perspective, routines for the control and monitoring of the opto-electronic peripherals such as the display, the laser and the camera, have been developed. Additionally, we have developed software for implementing the RBM key extraction along with the BCH error correction code and the whole enrolment and authentication protocol described in sections 2.2 and 2.3. All these routines are harmonically cooperating and organized by a main server platform that runs at the background managing the requests from external devices.

**4.4 Performance evaluation**

Up to now we have mentioned and experimentally demonstrated that the photonic PUF can be used as an authentication mechanism and as a random number generator. Using the properties and the metrics presented in sections 2.2 and 2.4, respectively and following a similar analysis as the one demonstrated in section 2.4 we are going to evaluate the performance of the developed device for the aforementioned functionalities.

For the authentication functionality, the first and most crucial property is the robustness of the device. This means that the device must reproduce the same CRPs for a long-term operation within a wide range of ambient temperatures. For this measurement, the PUF case is placed inside a closed room with an air-condition regulating the ambient temperature. The correlation diagram of the output speckle pattern generated by a specific randomly selected challenge was measured as a function of room temperature. The system was tested with and without the internal temperature stabilization mechanism, for a $5^0C$ range of temperature variations. The initial temperature of the room was $25^0C$ and followed a $+3^0C$ and a $-2^0C$ perturbation.

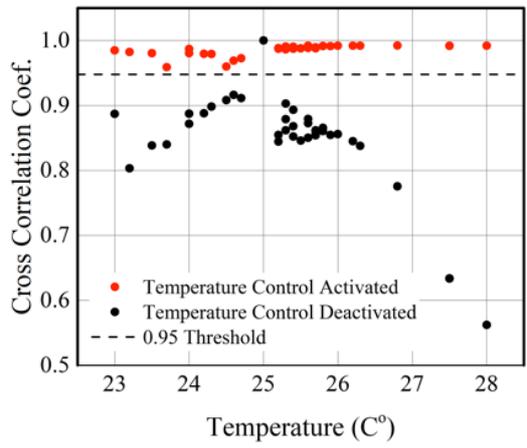

Figure 10. Single speckle pattern correlation as a function of temperature and time with the internal temperature control mechanism activated (red circles) and de-activated (blue squares).

As can be seen in figure 16, with the internal temperature control de-activated, the correlation is significantly degraded following the ambient temperature fluctuations. On the contrary, when the temperature control is activated, the correlation remains stable with a correlation above 0.95 during the entire measurement. Using a BCH code with error correction capability of 51 bits for a key length equal to 512, the errors generated by this amount of correlation can be fully compensated.

In order to further test the capabilities of the device in a long-term operation, an additional robustness measurement was conducted with the internal temperature control activated for several days. The temperature of the room during the measurement was $25^0$C with a deviation of 2 degrees. The results, as figure 17 clearly demonstrates, are rather encouraging proving that the specific device can offer long-term stability, which can be further enhanced through a better mechanical and thermal design.

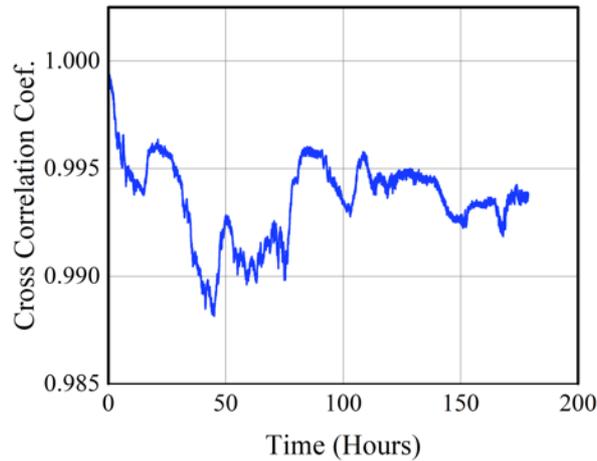

Figure 11. Long-term robustness with the internal temperature control activated

The next significant property of the device is its capability to recover after a power failure. By recovery we mean that the same CRPs can be regenerated even after the device is powered down. For this measurement, using the same challenge, the time elapsed for the correlation of the corresponding response to reach 0.99 was measured. The device was turned off for several days and the measurement started when the device was turned on again. As can be seen in the figure below, the time that is needed for the system to recover is only a few minutes. Obviously, this time depends on the ambient temperature, the reference temperature of the system and the speed of the temperature control mechanism. The reduction of the recovery time can be realized by selecting an optimal reference temperature for the system, and by enhancing the temperature stabilization mechanism.

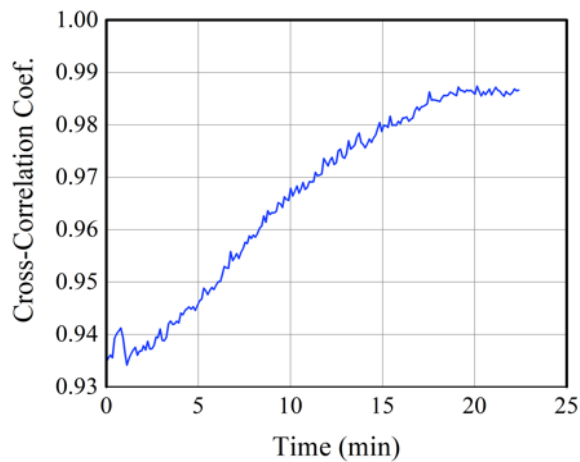

Figure 12. Recovery capability using the same CRP

The next measurement that must take place in order to fully characterize the capability of the specific device to operate as an authentication mechanism is the unpredictability, which defines the number of unique CRPs that can be generated by the system. As mentioned in section 4.1.2, the effective pixel area of the display that is lit by the laser beam is 350x350. This gives an enormous space of challenges. However, there is a minimum number of overlapping pixels between different challenges after which the responses become correlated. By excluding these challenges, the CRP space is reduced. Unfortunately, the theoretical calculation of such a combinatorics problem is impossible and time-consuming simulations have to be performed in order to estimate the CRP space. As an alternative, we performed an unpredictability measurement using randomly generated challenges and calculate the correlation between the corresponding responses. Obviously, this kind of measurement is practically impossible due to the large challenge space, however for a period of one week we manage to collect data and calculate the correlation of 5226 unique challenges. The probability distribution of the correlation between the corresponding responses is depicted in the figure below.

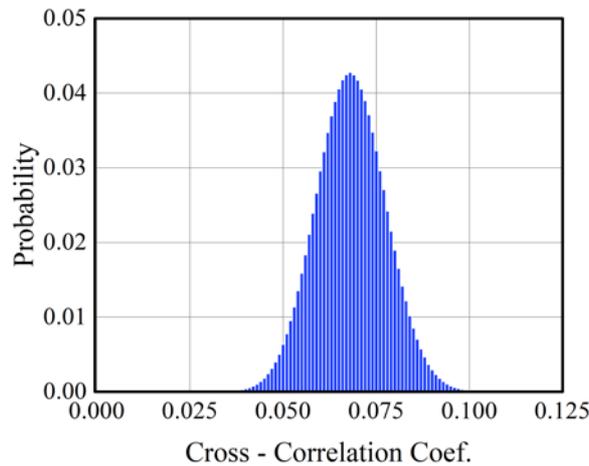
Figure 13. Unpredictability measurement using 5226 different challenges

In order to finalize the evaluation of the system as an authentication mechanism, we have to measure the unclonability of the system. For this purpose, we measured 250 different diffusers using the same illumination conditions. In the figure below, the probability distribution of the correlation between the response of each token using the same challenge is depicted.

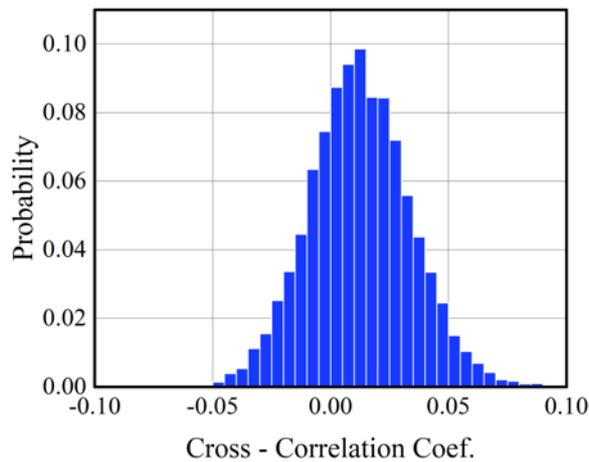
Figure 14 Unclonability measurement using 250 different PUF tokens

Finally, in order to evaluate the device as a random number generator, the NIST Statistical Test Suite (STS) was used. To construct the appropriate dataset defined by the NIST STS, the 5226 responses generated during the previous measurement were utilised. The images were processed through the RBM and from each one 20000 bits were extracted resulting in the acquisition of 104520000 bits in total, distributed in 100 *bitstreams with n* = 1045200 bitstream length.

The output of the NIST tests for the employed PUF module are presented in the following table. In order to acquire useful results, we used the same data in 3 different runs of the NIST STS, using all available tests. In the table, the worst results are presented. Also, in cases where multiple tests are run from the NIST STS, by default the worst results are presented. Each row in the table below corresponds to one test. Values in the "p-value uniformity" column represent the results (in the form of p-value) for uniformity testing of p-values computed for a given test. Value in the column "Proportion" represents proportion of sequences that pass a given test.

Table 1. The results of the NIST STS. "p-value" corresponds to the uniformity of the results whereas "Proportion" corresponds to the percentage of bit strings that passed the test.

| STATISTICAL TEST | PROPORTION | P-VALUE UNIFORMITY |
|---|---|---|
| **Frequency** | 100/100 | 0.000145 |
| **Block Frequency** | 99/100 | 0.066882 |
| **Cumulative Sums** | 100/100 | 0.000000 |
| **Runs** | 99/100 | 0.514124 |
| **Longest Run** | 99/100 | 0.719747 |
| **Rank** | 99/100 | 0.574903 |
| **FFT** | 99/100 | 0.514124 |
| **Non-Overlapping Template** | 99/100 | 0.779188 |
| **Overlapping Template** | 98/100 | 0.867692 |
| **Universal** | 99/100 | 0.924076 |
| **Approximate Entropy** | 99/100 | 0.554420 |
| **Random Excursions** | 67/68 | 0.253551 |
| **Random Excursions Variant** | 67/68 | 0.437274 |
| **Serial** | 98/100 | 0.883171 |
| **Linear Complexity** | 100/100 | 0.534146 |

The data sequences generated by our system, as illustrated in **Table 1**, have succeeded in all "proportion" tests and in the vast majority of "p-value uniformity" tests of the NIST suite (14 / 15 passed), only failing to pass the uniformity testing of p-values for the "Cumulative Sums" test, which is one of the 3 tests that evaluates the percentage of "zeros" and "ones" in the bit sequences. As can be seen from the detailed p-values for this test that are presented in the figure below, this test does not produce marginal results (low p-values), but, on the contrary, success rate is too high, which implies a highly balanced percentage of "zeros" and "ones" in the bit sequences. Nonetheless, this is a statistical anomaly which does not impose a security breach, due to the fact that an adversary has no indication regarding the probability of bit-flips. Therefore, the proposed PUF can be considered as an adequate random number generator.

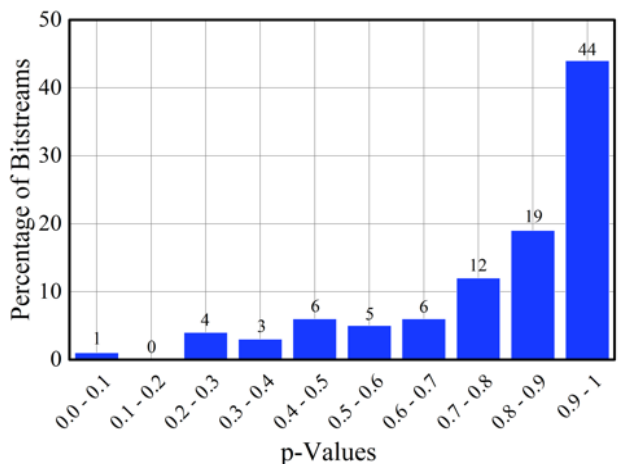

Figure 15. Statistical analysis of p-values

## 5. CONCLUSION

In this work, the complete process from conceptual idea to the implementation of a standalone, fully working and miniaturized PUF device was demonstrated. During this process, a full theoretical investigation on photonic PUFs was conducted, followed by an extensive experimental analysis on the properties and capabilities of different PUF tokens. The advantage of a diffuser as a PUF token using an intensity pattern as challenge generator was demonstrated leading to the selection of the appropriate components for the implementation of a standalone PUF device. The final device demonstrated enhanced performance in terms of robustness, maintaining its stability for temperature variations of $5^0$C and during continuous operation over a period of 1 week. Additionally, the system demonstrated exponentially large CRP space, negligible probability of hardware replication and capability to fully recover after a power failure. Up to our knowledge, this is the first time that a standalone and compact photonic PUF device operating in a simulated environment is presented in the literature. The system proved its capability to function as a random number generator and as an authentication mechanism and through minor optimizations its performance and miniaturization level can be further improved.

## ACKNOWLEDGEMENTS

The research leading to these results has received funding from the European Union's Horizon 2020 research and innovation programme under grant agreement No 740931 (SMILE – Smart Mobility at the European Land Borders), and part of this work was implemented by scholarship from the IKY (State Scholarships Foundation) act "Reinforcement of research potential through doctoral research" funded by the Operational Programme "Human Resources Development, Education and Lifelong Learning", 2014-2020, co-financed by the European Commission (European Social Fund) and Greek state funds. This work was also co-funded by the European Union and Greek national funds through the Operational Program "Competitiveness, Entrepreneurship and Innovation", under the call "Strengthening Research and Innovation
Infrastructures" (project code MIS: 5002735).

# REFERENCES

[1] "Big data analytics | IBM Analytics." [Online]. Available: https://www-01.ibm.com/software/data/ bigdata/what-is- big- data.html.

[2] R. Pappu, B. Recht, J. Taylor, and N. Gershenfeld, "Physical One-Way Functions," *Science (80-. ).*, vol. 297, no. 5589, pp. 2026–2030, 2002.

[3] D. Lim, J. W. Lee, B. Gassend, E. G. Suh, M. van Dijk, and S. Devadas, "Extracting secret keys from integrated circuits.," in *Very Large Scale Integration (VLSI) Systems, IEEE Transactions*, 2005, vol. 13, no. 10, pp. 1081–1085.

[4] M. D. M. Yu, R. Sowell, A. Singh, D. M'Raihi, and S. Devadas, "Performance metrics and empirical results of a PUF cryptographic key generation ASIC," in *Proc. of the IEEE Int. Symposium on Hardware-Oriented Security and Trust (HOST)*, 2012, pp. 108–115.

[5] S. S. Kumar, J. Guajardo, R. Maes, G. J. Schrijen, and P. Tuyls, "The Butterfly PUF protecting IP on every FPGA," in *Proc. of the IEEE International Workshop on Hardware-Oriented Security and Trust (HOST)*, 2008, no. 71369, pp. 67–70.

[6] A.-R. Sadeghi, I. Visconti, and C. Wachsmann, "Enhancing RFID Security and Privacy by Physically Unclonable Functions," in *Towards Hardware-Intrinsic Security: Foundations and Practice*, A.-R. Sadeghi and D. Naccache, Eds. Springer, 2010, pp. 281–305.

[7] S. Devadas, E. Suh, S. Paral, R. Sowell, T. Ziola, and V. Khandelwal, "Design and Implementation of PUF-Based 'Unclonable' RFID ICs for Anti-Counterfeiting and Security Applications," in *Proc. of the IEEE International Conference on RFID*, 2008, pp. 58–64.

[8] P. Qiu, Y. Lyu, J. Zhang, X. Wang, D. Zhai, D. Wang, and G. Qu, "Physical Unclonable Functions-based Linear Encryption against Code Reuse Attacks.," in *Proc. of the 53rd Annu. Conf. Des. Autom.*, 2016.

[9] B. Gassend, D. Clarke, M. van Dijk, and S. Devadas, "Silicon physical random functions," in *Proc. of the 9th ACM Conference on Computer and Communications Security*, 2002, no. November, pp. 148–160.

[10] J. W. Lee, D. L. D. Lim, B. Gassend, G. E. Suh, M. Van Dijk, and S. Devadas, "A technique to build a secret key in integrated circuits for identification and authentication applications," in *Proc. of the IEEE Symposium on VLSI Circuits. Digest of Technical Papers*, 2004, pp. 176–179.

[11] D. E. Holcomb, W. P. Burleson, and K. Fu, "Power-Up SRAM state as an identifying fingerprint and source of true random numbers," *IEEE Trans. Comput.*, vol. 58, no. 9, pp. 1198–1210, 2009.

[12] R. Liu, H. Wu, Y. Pang, H. Qian, and S. Yu, "Experimental Characterization of Physical Unclonable Function Based on 1kb Resistive Random Access Memory Arrays," *IEEE Electron Device Lett.*, vol. 36, no. 12, pp. 1380–1383, 2015.

[13] G. T. Becker, "The Gap Between Promise and Reality : On the Insecurity of XOR Arbiter PUFs," in *Proc. of the 17th Int. Workshop on Cryptographic Hardware and Embedded Systems*, 2015, pp. 535–555.

[14] Y. Cao, L. Zhang, S. S. Zalivaka, C.-H. Chang, and S. Chen, "CMOS Image Sensor Based Physical Unclonable Function for Coherent Sensor-Level Authentication," *IEEE Trans. Circuits Syst. I Regul. Pap.*, vol. 62, no. 11, pp. 2629–2640, 2015.

[15] P. H. Nguyen, D. P. Sahoo, R. S. Chakraborty, and D. Mukhopadhyay, "Efficient Attacks on Robust Ring Oscillator PUF with Enhanced Challenge-Response Set," in *Proc. of the Design, Automation & Test in Europe Conference & Exhibition (DATE)*, 2015, pp. 641–646.

[16] G. Hospodar, R. Maes, and I. Verbauwhede, "Machine learning attacks on 65nm Arbiter PUFs: Accurate modeling poses strict bounds on usability," in *Proc. of the 2012 IEEE International Workshop on Information Forensics and Security (WIFS)*, 2012, pp. 37–42.

[17] U. Rührmair, F. Sehnke, J. S ölter, G. Dror, S. Devadas, and J. Ü. Schmidhuber, "Modeling attacks on physical unclonable functions," in *Proc. of the 17th ACM conference on Computer and communications security - CCS '10*, 2010, p. 237.

[18] S. Tajik, F. Ganji, J. P. Seifert, H. Lohrke, and C. Boit, "Laser fault attack on physically unclonable functions," in *Proc. of 2015 Workshop on Fault Diagnosis and Tolerance in Cryptography (FDTC)*, 2016, pp. 85–96.

[19] A. Mahmoud, U. Rührmair, M. Majzoobi, and F. Koushanfar, "Combined Modeling and Side Channel Attacks on Strong PUFs," 2013.

[20] U. Rührmair, X. Xu, J. Sölter, A. Mahmoud, M. Majzoobi, F. Koushanfar, and W. Burleson, "Efficient Power and Timing Side Channels for Physical Unclonable Functions," *Cryptogr. Hardw. Embed. Syst.*, no. 8731, pp. 476–492, 2014.

[21] U. Rührmair, S. Urban, A. Weiershäuser, and B. Forster, "Revisiting Optical Physical Unclonable Functions," *Cryptol. ePrint Arch.*, vol. 215, pp. 1–11, 2013.

[22] H. Zhang and S. Tzortzakis, "Robust authentication through stochastic femtosecond laser filament induced scattering surfaces," *Appl. Phys. Lett.*, vol. 108, p. 211107, 2016.

[23] S. Shariati, F.-X. Standaert, L. Jacques, and B. Macq, "Analysis and experimental evaluation of image-based PUFs," *J. Cryptogr. Eng.*, vol. 2, no. 3, pp. 189–206, 2012.

[24] J. D. R. Buchanan, R. P. Cowburn, A.-V. Jausovec, D. Petit, P. Seem, G. Xiong, D. Atkinson, K. Fenton, D. a Allwood, and M. T. Bryan, "Forgery: 'fingerprinting' documents and packaging.," *Nature*, vol. 436, no. 7050, p. 475, 2005.

[25] C. Mesaritakis, M. Akriotou, A. Kapsalis, E. Grivas, C. Chaintoutis, T. Nikas, and D. Syvridis, "Physical Unclonable Function based on a Multi-Mode Optical Waveguide," *Sci. Rep.*, vol. 8, 2018.

[26] M. Akriotou, C. Mesaritakis, and E. Grivas, "Random Number Generation from a Secure Photonic Physical Unclonable Hardware Module," in *Security in Computer and Information Sciences. Euro-CYBERSEC 2018. Communications in Computer and Information Science*, vol. 821, Springer International Publishing, 2018, pp. 28–37.

[27] U. Rührmair, C. Hilgers, and S. Urban, "Optical PUFs Reloaded," *Eprint.Iacr.Org*, 2013.

[28] U. Maurer, R. Renner, and S. Wolf, *Security with Noisy Data: On Private Biometrics, Secure Key Storage and Anti-Counterfeiting.* Springer Science & Business Media, 2007.

[29] Y. Dodis, R. Ostrovsky, L. Reyzin, and A. Smith, "Fuzzy Extractors : How to Generate Strong Keys from Biometrics and Other Noisy Data," *SIAM J. Comput.*, vol. 38, no. 1, pp. 97–139, 2008.

[30] F. Armknecht, R. Maes, A. R. Sadeghi, F. X. Standaert, and C. Wachsmann, "A formal foundation for the security features of physical functions," *Proc. - IEEE Symp. Secur. Priv.*, pp. 397–412, 2011.

[31] R. A. Sadek, "SVD Based Image Processing Applications : State of The Art , Contributions and Research Challenges," *Int. J. Adv. Comput. Sci. Appl.*, vol. 3, no. 7, pp. 26–34, 2012.

[32] S. S. Kozat, R. Venkatesan, and M. K. Mihcak, "Robust perceptual image hashing via matrix invariants," in *International Conference on Image Processing (ICIP)*, 2004, pp. 3443–3446.

[33] E. J. Candès and M. B. Wakin, "An Introduction To Compressive Sampling," *IEEE Signal Process. Mag.*, vol. 25, no. 2, pp. 21–30, 2008.